\documentclass[twocolumn,superscriptaddress,amsmath,amssymb,floatfix,aps,notitlepage,nokeys,prl,longbibliography,nofootinbib,longbibliography,aps,prd]{revtex4-2}
\usepackage{mathrsfs, hyperref}
\usepackage{amssymb, amsbsy, amsmath, latexsym, dsfont, array, layout, graphics,mathrsfs,braket,amsfonts,amsthm,
  amssymb,graphicx,subfigure,dcolumn, youngtab,color,bm,mathtools,braket,verbatim,url,cleveref,natbib,hypernat,extarrows,inputenc,multirow}
\usepackage{lipsum}

\newcommand{\sg}[1]{{\color{black} #1}} 
\usepackage[mathlines]{lineno}

\newcommand{\red}[1]{\textcolor{black}{#1}}

\begin{document}
\preprint{APS/123-QED}

\title{Centipede-Like Metastrip Enables On-Demand Programmable Drop Motion} 

\author{Sergio Britto}
\affiliation{%
Department of Civil, Environmental, and Geo- Engineering,
University of Minnesota, Minneapolis, MN 55455, US}

\author{Stefano Gonella}%
\email{sgonella@umn.edu}
\affiliation{%
Department of Civil, Environmental, and Geo- Engineering,
University of Minnesota, Minneapolis, MN 55455, US}%


\begin{abstract}


We demonstrate 
programmable, frequency-tunable and size-selective drop motion 
on an elastic metastrip substrate with a centipede-like array of cantilever resonators.
Under harmonic excitation, the strip experiences simultaneously an in-plane (IP) collective motion and out-of-plane (OOP) deformation, 
thus establishing a frequency-dependent IP–OOP phase landscape. This prescribes the direction of motion to the drops on the strip based on their position, 
causing the emergence of 
clustering and rarefaction regions. 
By tuning the 
tip masses of the resonators, 
we open reconfigurable OOP bandgaps that modify the IP-OOP phase makeup and locally suppress drop motion, contributing an additional layer of spatial selectivity. 
Using multi-frequency excitations, we 
selectively actuate drops based on their resonances, 
effectively filtering them by volume and position. 

\end{abstract}
\maketitle


Drop motion mechanics govern many natural and artificial processes, from rain formation~\cite{Khain2013} and printing methods~\cite{Wijshoff2018} to 
microfluidic systems~\cite{shesto2004, Seemann2012, Stone2004, Squires2005}, energy harvesting~\cite{Mugele2005, Li2022} and biological phenomena~\cite{Blossey2002, Yu2019, Zwicker2025, Zeng2023}. The ability to control drop motion on surfaces has been achieved through a range of 
chemical, physical and mechanical
processes~\cite{Daniel2002, Daniel2004, Daniel2005, Cira-et-al_Drop-Motility_Nature_2015, Mrinal-et-al_Ratchet-Leidenfrost_Langmuir_2017, Graeber-et-al_Leidenfrost-Trampoline_Nat-Comm_2021, Hartmann2022, Li-et-al_Leidenfrost-Tailoring_Nat-Comm_2023}. Several studies have targeted 
drop motion via surface vibration~\cite{Celestini2006, Dong2006, Brunet2007, Brunet2009, Noblin2009, John2010, Whitehill2010, Sartori2015, Costalonga2020, Deegan2020}. This effect arises from the activation and 
superposition of sessile drop modes~\cite{Steen2019}, which generate asymmetric capillary forces within each excitation cycle, producing a net force that overcomes the drop pinning forces~\cite{Butt2022}. 
Achieving more complex effects, \textit{i.e.}, individually actuating selected drops based on their position, requires moving beyond the canonical paradigm of rigid substrates. Using \textit{elastic} substrates, one can leverage their deformability to achieve a spatially variable vibrational landscape and elicit spatially selective drop activation conditions. 

The potential of elastic substrates for spatially selective drop control was explored in~\cite{Charara2025} using a metaplate as vibrating substrate: populating a region with an array of 
pillars would open a locally resonant bandgap within a specific frequency range, locally attenuating the plate response. This mechanism was exploited to pattern the motion of sliding drops on an inclined surface, where the drops would come to a stop 
upon reaching the attenuation zone. Although this approach enables programming depinning and 
sliding, other regimes - such as net motion on horizontal surfaces or climbing - remain out of reach. Achieving such conditions requires the substrate to undergo substantial displacements both in-plane (IP) and out-of-plane (OOP) to activate \textit{simultaneously} the drop \textit{rocking} and \textit{pumping} modes, the combination of which (under appropriate phase conditions) is a necessary ingredient for net 
motion~\cite{Brunet2007, Noblin2009}. However, plates are much stiffer IP than OOP, resulting in a negligible IP response. Therefore, to fully harness the potential of elastic substrates, it is necessary to design alternative structures that display \textit{comparable} IP and OOP compliance.

In prior work~\cite{Britto2025}, we addressed this need by introducing a substrate design consisting of a plate element supported by thin elastic ligaments at which the deformation localizes. Under harmonic 
excitation, the substrate could 
undergo simultaneously IP and OOP 
displacements, with the appropriate phase 
requirements, \textit{i.e.}, the elastic response automatically tailored the established 
phase landscape to the excitation frequency. This configuration allowed achieving net drop motion and 
climbing using a single actuation source, as well as selecting the drop motion direction by simply varying the excitation frequency. Formally, we were able to link the different motion regimes to a relative IP-OOP phase metric $\Delta \varphi (f)$, with $f$ denoting frequency. Using $\Delta \varphi$, we could also forecast the collective motion of multiple drops deposited on multi-cell configurations. Despite these perks, 
the discreteness of multi-cell configurations imposes a drastic constraint on drop mobility, limiting programmability to finite domains, penalizing the versatility of the approach. 

In this Letter, we definitively overcome this limitation by conceiving a type of substrate that preserves - and elevates - the attributes of IP-OOP compliance 
while deploying a \textit{continuous} domain for unconstrained drop motion planning. Moreover, the proposed substrate deploys a metamaterial architecture with rich tunability features, which enables a host of spectro-spatial corrections of the vibrational response and imparts additional spatial selectivity to the drop motion landscape. The approach 
blends notions from 
metamaterials engineering, fluid mechanics and locomotion robophysics. On the one hand, by studying  metamaterial-enabled surface-drop dynamics, we add one layer to the burgeoning 
panorama of fluid-structure interactions enabled by phononic materials, which includes recent applications to flow control~\cite{Hussein-et-al_Stabilization-Phononics_PRSA_2015,Avalone_Meta-Flow-Review_arXiv_2025, Armin-Hussein_Flow-Control_JAP_2023}. On the other hand, we transfer to a fluids problem some paradigms 
developed for vibration-based locomotion or self-assembly of solid objects~\cite{Kudrolli_Self-Propelled-Rods_PRL_2010,Aguilar-et-al_Locomotion-Robophysics-Review_Rep-Prog-Phys_2016,Barois-et-al_Locomotion-Stick-Slip_PRL_2024,Watkins-Bilal_Reprogramamble-Magnets_PRApp_2025}.

\begin{figure}[t!]
\includegraphics[width=\columnwidth]{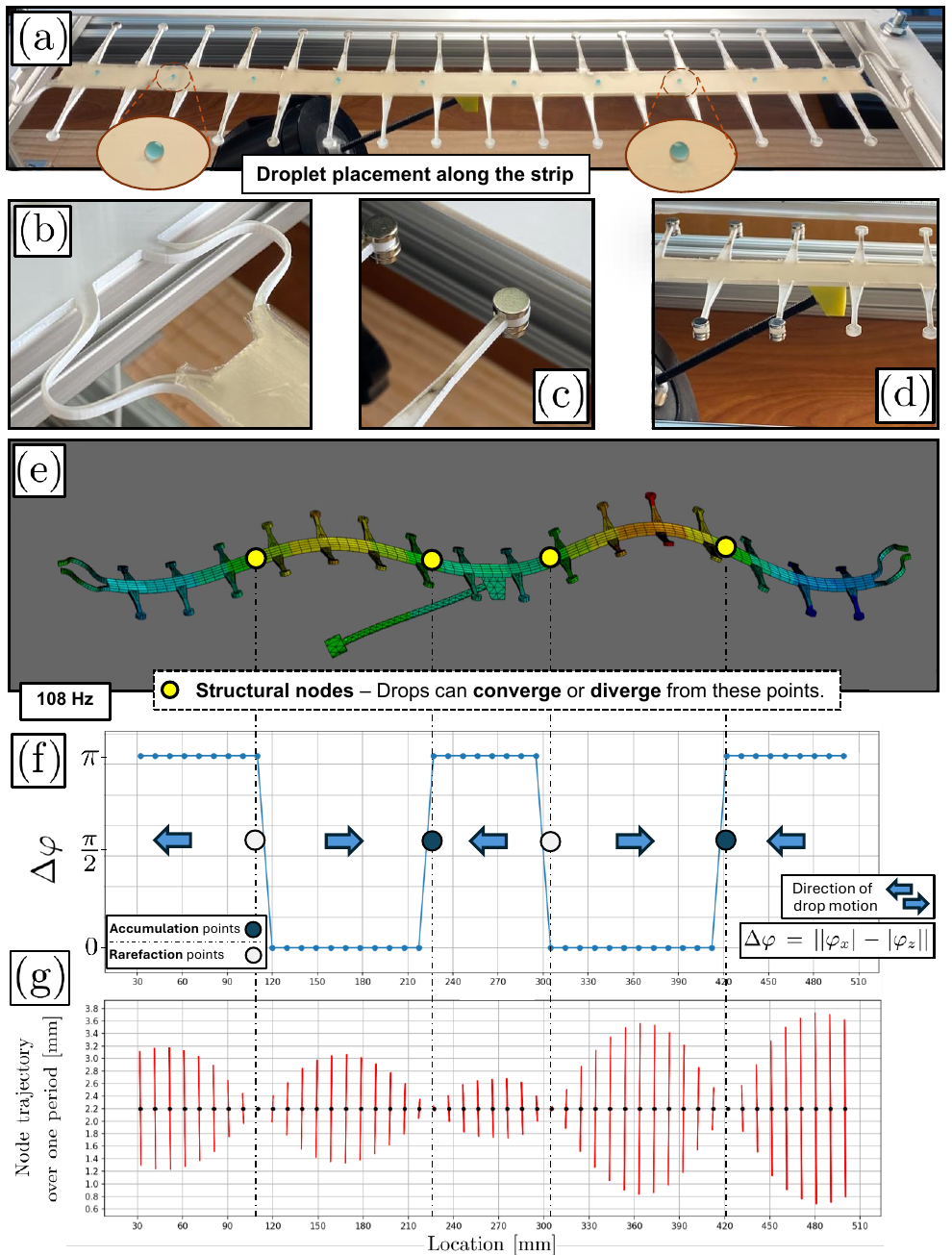}
\caption{\label{fig:f1} Prototype of metastrip for drop motion control. 
(a) 
Strip mounted on frame and excited via a shaker, with drops deposited on the substrate. 
(b) Close-up of curved elastic ligaments providing in-plane compliance.
(c) Lateral cantilever resonators with variable tip masses for tunability.
(d) Details of wedge and stinger.
(e) ANSYS finite element (FE) simulation at 108 Hz and 5 N excitation amplitude showing regularly-spaced structural nodes (yellow markers).
(f) Relative IP-OOP phase metric $\Delta \varphi$, plotted along the strip span, predicting the direction of expected droplet motion (blue arrows) and denoting structural nodes as either accumulation or rarefaction points.
(g) Simulated spatial trajectories of points along the strip over one period, displaying change of inclination across the structural nodes.
}
\end{figure}

We envision our substrate as a continuous strip (Fig.~\ref{fig:f1}(a)), 
supported by pairs of curved structural ligaments detailed in Fig.~\ref{fig:f1}(b). Following the design principles introduced for discrete substrates in~\cite{Britto2025}, here the IP compliance is primarily provided by the ligaments, which can undergo significant localized bending deformation, whereas the 
OOP compliance arises naturally from the 
inherent bending deformability of the strip. 
Furthermore, we equip the strip with two arrays of equally-spaced lateral cantilevers with tip masses. This produces two effects: 1) it renders the strip a periodic structure; 2) it equips each emergent unit cell with a pair of beam-like resonators, shown in Fig.~\ref{fig:f1}(c), which can be activated in selected frequency ranges to enable additional control over the substrate’s vibrational response. 
The tip masses are adjustable via stackable magnetic discs to tune the resonators natural frequencies and, consequently, the strip vibrational response. These features render the structure a 1D locally-resonant metastrip~\cite{Widstrand2022} in which 
resonances govern bandgap formation. At low frequencies, the resonators remain largely unengaged, bearing a negligible effect on the strip dynamics - besides a small correction of its inertial properties. In contrast, at higher frequencies that approach their resonance, they are strongly activated, opening a locally-resonant bandgap and attenuating the vibration of the strip segment to which they are attached.


The strip is fabricated by laser cutting a thin acrylic sheet (Young's modulus $E = 3.3 \, \textrm{GPa}$, Poisson's ratio $\nu = 0.35$, thickness $2.2 \, \textrm{mm}$). It measures $532 \, \textrm{mm}$ in length and is anchored to a frame through the ligaments. A wedge is attached to the bottom surface (Fig.~\ref{fig:f1}(d)) to apply a force excitation at a prescribed inclination (here $30^{\circ}$ from the horizontal) via the stinger of a shaker (Br\"{u}el \& Kj\ae r Type 4810 Mini). \textcolor{black}{We ensure consistent hydrophobic wetting behavior and hysteresis by applying a low-friction PTFE (polytetrafluoroethylene)} tape to the top surface, providing an equilibrium contact angle of $94^{\circ} \pm 3^{\circ}$, an advancing angle of $92^{\circ} \pm 4^{\circ}$, and a receding angle of $73^{\circ} \pm 3^{\circ}$. \textcolor{black}{To guarantee a consistent and meaningful characterization of the system, we enforce that the droplet operates under the same modal conditions across all excitation frequencies. Specifically, the droplet response is maintained in the range bounded below and above by the resonant frequencies of its rocking and pumping mode, respectively, for every case considered, following considerations in~\cite{Noblin2009, Brunet2007}. To achieve this, the droplet size is adjusted from case to case: for the rocking mode according to the semi-analytical sessile drop eigenfrequency criteria proposed in~\cite{Celestini2006}, alongside the spherical cap volume calculation in~\cite{Sartori2015}; for the pumping mode following the qualitative argument proposed in~\cite{Noblin2009}.} 



\begin{figure}[t!]
\includegraphics[width=\columnwidth]{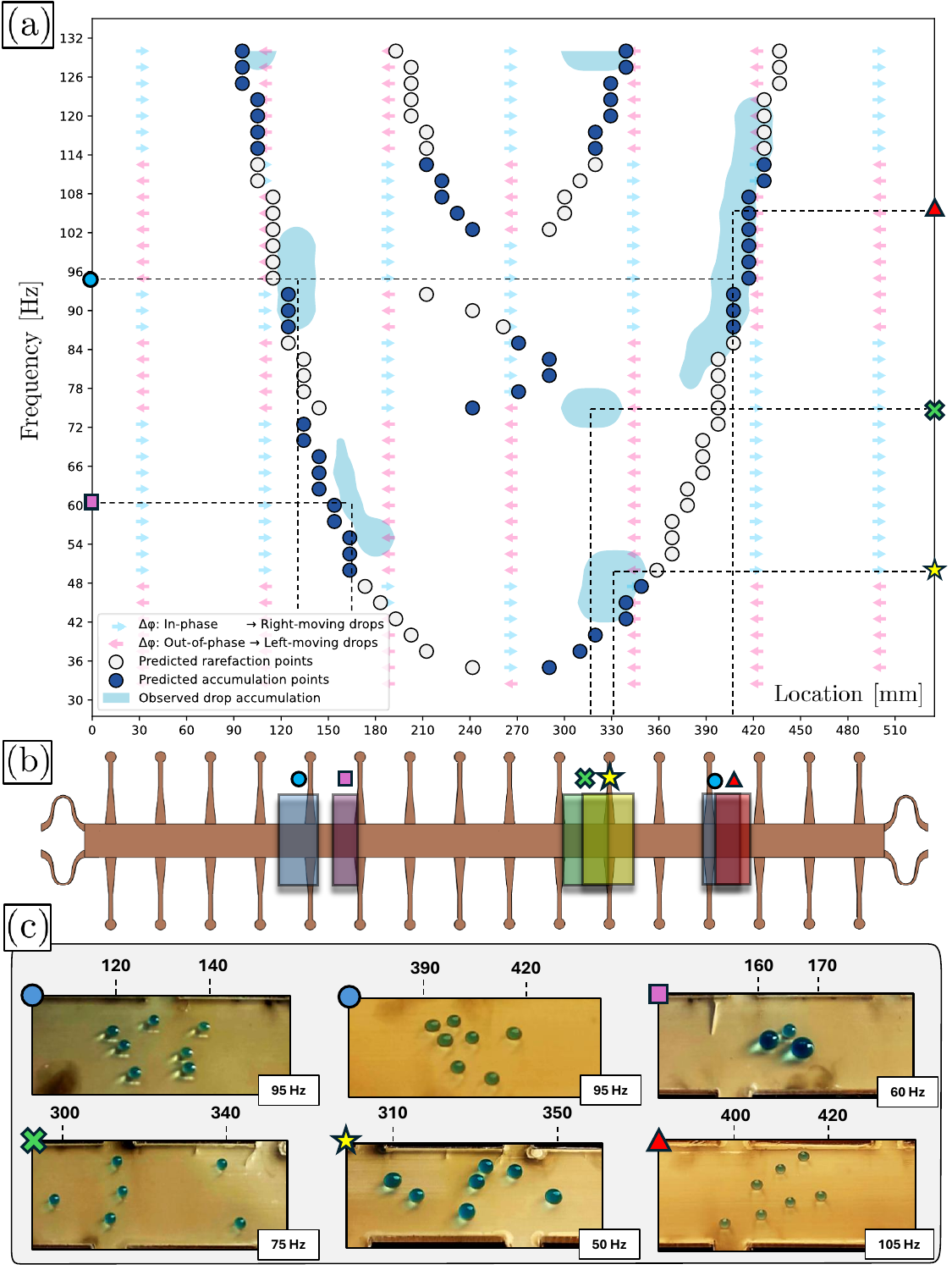}
\caption{\label{fig:f2} (a) Predicted and experimentally verified accumulation/rarefaction points for different excitation frequencies. 
Blue (pink) arrows denote predicted right (left) 
motion based on $\Delta \varphi$ values. Blue (gray) dots mark 
predicted accumulation (rarefaction) points. Experimentally observed accumulation regions (blue shaded regions) are overlaid. (b-c) Snapshots of drop clusters observed upon vibration at selected frequencies color-coded in (a), with clusters positions indicated along the strip. \sg{Details on excitation amplitude in SM.}}
\end{figure}


\begin{figure*}[t!]
\includegraphics[width=\textwidth]{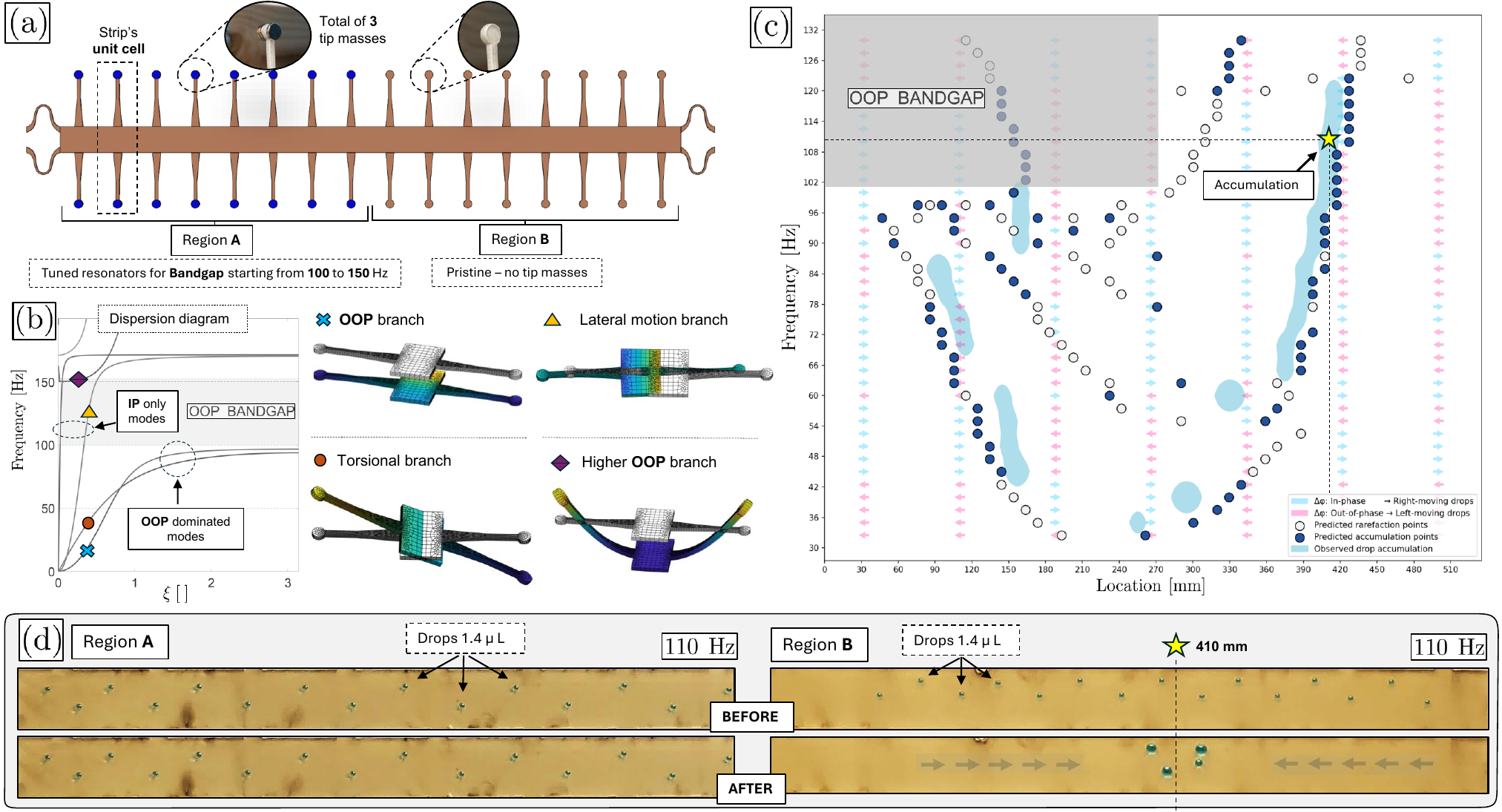}
\caption{\label{fig:f3} Effect of OOP bandgap on drop motion.
(a) Schematic of the strip: Region A with activated (mass-augmented) resonators and Region B pristine. 
(b) Band diagram of strip from Bloch analysis, revealing OOP modal bandgap between 100 and 150 Hz, where OOP strip response and drop motion are suppressed.
(c) Frequency with location and frequency range of bandgap marked. 
Attenuated substrate motion suppresses drop motion locally, reflected by the lack of experimentally observed drop accumulation clusters.
(d) Drop motion snapshots (before and after vibration) at 
110 Hz (within the bandgap), in Region A and B. In A substrate response attenuation causes drops to remain stationary, while in B drops move and accumulate at the predicted location (bigger drops on the image are clusters of the smaller drops upon accumulation).}
\end{figure*}

We start our analysis with a pristine configuration without extra tip masses on the resonators. Here, 
the effects of the resonators are negligible, since their natural frequencies lie well above our frequency range of interest. 
Therefore, the strip can be treated effectively as a uniform beam deforming in bending, and any spatial heterogeneity of the response along its span comes solely from the operational deflection shapes 
expected in the considered frequency range. Finite element (FE) simulation of the harmonic response at a representative excitation frequency (108 Hz), shown in Fig.~\ref{fig:f1}(e), reveals a deflection pattern exhibiting structural nodes at which the OOP displacement vanishes. Across a structural node, the response undergoes a phase change, \textit{i.e.} points located to the left and to the right move OOP in opposition of phase. In contrast, IP motion is entirely controlled by the curved ligaments, where the IP deformation localizes. Therefore, at this frequency, it is effectively in-phase for all points along the strip,
\textit{i.e.}, the IP motion can be treated as a rigid-body motion (details in supplemental material (SM)). 
As a result, at the structural nodes, the IP and OOP responses undergo jumps in \textit{relative phase}. To capture this, 
we invoke the relative IP-OOP phase metric $\Delta \varphi$, previously used to characterize vibrations of discrete chains~\cite{Britto2025}. The curve of $\Delta \varphi$ as a function of position $x$ at this frequency (Fig.~\ref{fig:f1}(f)) indeed features $\pi$ jumps at the structural node locations: 110, 230, 330, and 420 mm. In Fig.~\ref{fig:f1}(g) we plot the spatial trajectories (over one 
cycle) of material points sampled along the strip. The trend confirms the vanishing motion at the nodes and the 
changes in trajectory inclination that occur across them corroborate the phase shifts. 

Any drop moving on the strip is expected to come to a rest upon reaching a structural node. To predict the behavior of drops in regions separated by a node, it is sufficient to inspect the values of $\Delta \varphi$ across the node. Invoking the phase-drop motion relations discussed in~\cite{Britto2025}, we expect a net drop displacement to the right (R) 
in regions where $\Delta \varphi=0$ and 
a net drop motion to the left (L) 
in regions where $\Delta \varphi=\pi$, as indicated by the arrows in Fig.~\ref{fig:f1}(f). 
The sequence/switch of motion directions observed across a node (R $\rightarrow$ L vs. L $\rightarrow$ R) allows classifying the structural node as either an \textit{accumulation} point -- exhibiting convergent drop motion -- or a \textit{rarefaction} point -- exhibiting divergent drop motion, respectively. Compared to the case of discrete substrates, the higher modal richness of continua  induces a proliferation of nodes and possible accumulation/rarefaction points.

This analysis can be replicated over a broad interval of frequencies. Sweeping frequency, we expect a variety of regimes to emerge, governed by the number and placement of structural nodes along the strip, which are a function of frequency, and by their convergent or divergent character. In Fig.~\ref{fig:f2}, we compare structural nodes predictions, obtained from structural dynamics simulations, against the drop motion regimes actually observed experimentally in the 30 -- 132 Hz range. For each frequency, we mark the positions of the accumulation points (blue markers) and rarefaction points (gray) and we overlay the experimentally observed drop accumulation regions (shaded regions), reporting convincing agreement \sg{for most clustering scenarios.}
For a few representative frequencies (color-coded), we provide snapshots of the drop clusters observed upon vibration, which confirm accumulations at the predicted points (Movies 1--5 in SM). Note that, at higher frequencies (above 70 Hz) the OOP dynamics of the strip become more complex, 
with additional interior nodes emerging close to the mid-span and the outer nodes migrating towards the strip ends. Accordingly, additional accumulation points become, in principle, available. Our observations confirm the emergence of some 
of these new accumulation points (\textit{e.g.}, green marker at 75 Hz). 
\textcolor{black}{In some cases, we report shifts in the frequency ranges in which accumulation is observed (\textit{e.g.}, the upper bounds of the accumulation ranges $\approx$ 96 and 114 Hz) and/or in the location of the accumulation points. Such discrepancies are attributable to unavoidable differences between model and prototype (inaccuracies in laser-cut fabrication, material damping, effects of tape, etc.), which could cause imperfect predictions of the substrate stiffness. Other cases are not established in practice due to unfavorable IP-OOP amplitude ratios, which drive the pumping and rocking modes at unbalanced amplitudes, or to overall insufficient amplitude levels. Here, the IP and OOP displacement profiles are intrinsically dictated by the structural behavior of the substrate; as a result, while increasing the excitation amplitude can in some cases compensate for suboptimal conditions, there exist configurations where the magnitude mismatch between IP and OOP responses remains too unfavorable --or the response remains generally too weak-- to sustain drop motion. This is likely the case of the lack of accumulation points observed experimentally between $\approx$ 102 and 114 Hz, where the IP and OOP response remains locally too low (Fig. 1(d)  in SM).} 


Next we propose a configuration designed to prevent drop motion at selected locations via spatially-selective bandgap opening. To this end, we activate the resonators by populating the lateral beams with enough tip masses to lower their resonant frequencies down to our band of interest, as shown in Fig.~\ref{fig:f3}(a). Through 1D Bloch analysis, we compute the strip dispersion diagram, shown in Fig.~\ref{fig:f3}(b) with selected unit cell mode shapes to classify the modal character (IP vs. OOP) of the branches.  
The first two modes with some OOP character (red and blue \sg{markers}) fold at $\approx$ 100 Hz. The next mode with OOP character starts at $\approx$ 150 Hz, leaving a $\approx$ 50 Hz interval of \textit{modal OOP bandgap}, where OOP motion is forbidden and the \sg{only allowed phonons are IP (highlight in the figure)}.
We now subdivide the strip into two regions: Region A, with activated (mass-augmented) resonators, and Region B, with inactive (pristine) resonators. Exciting a frequency in the bandgap, drops in Region A are expected to remain stationary due to the attenuated OOP motion, whereas drops in Region B shall move and accumulate at the predicted locations. 
Fig.~\ref{fig:f3}(c) shows the full frequency sweep, with the 
bandgap region marked. 
Phase considerations suggest that accumulation points may still exist in the bandgap zone and frequency range, highlighted in figure, albeit with positions affected by the 
modified strip dynamics, but the 
response attenuation due to the bandgap trivializes their impact. 
In Fig.~\ref{fig:f3}(d), we show the results for a representative frequency of 110 Hz. In Region A the drops remain stationary even upon vibration, while in B they are set in motion and cluster at the predicted accumulation point (Movie 6 in SM). We conclude that resonators can be selectively tuned and placed to inhibit drop motion at 
\sg{desired} locations, granting an additional level of control.


\begin{figure}[t!]
\includegraphics[width=\columnwidth]{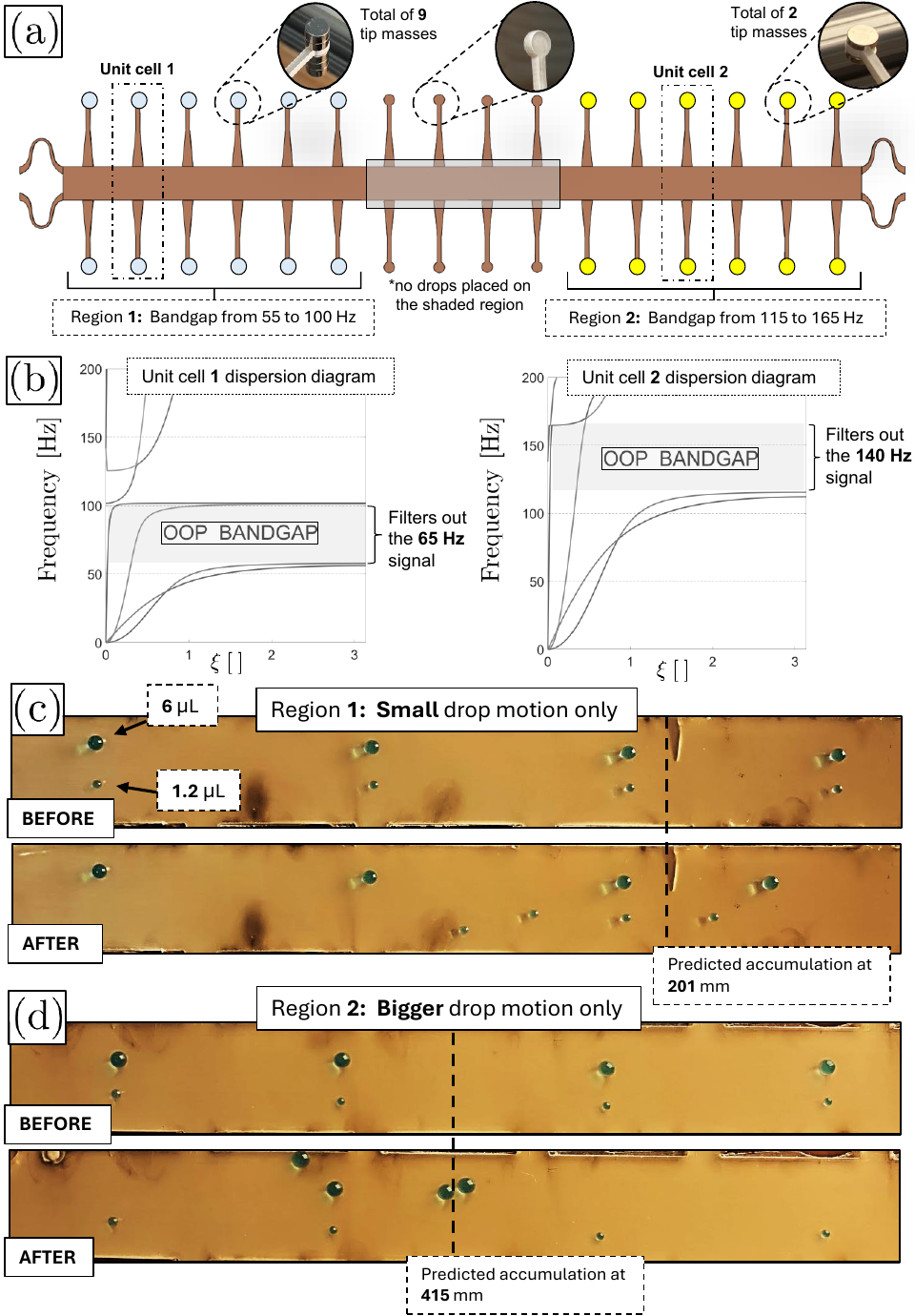}
\caption{\label{fig:f4} (a) Strip with regions with  beams resonating at different frequencies, acting as a spatial volume filter. (b) Band diagrams for Regions 1 and 2, featuring a low and high OOP bandgap, respectively. (c) Drop behavior in Region 1, with only small drops significantly moving and clustering. (d) Drop behavior in Region 2, with only large drops activated.
}
\end{figure}

We can stretch this concept to achieve even more complex spatial patterns and additional functionalities. Consider, \textit{e.g.}, the configuration in Fig.~\ref{fig:f4}(a), where we populate the beams in the first six cells (Region 1) with nine tip masses, and the last six cells (Region 2) with two tip masses. The dispersion diagrams for the two configurations, shown in Fig.~\ref{fig:f4}(b), reveal the opening of an OOP bandgap at a lower frequency in Region 1 and at a higher-frequency in Region 2. 
We prescribe a multi-frequency excitation, here the sum of two harmonic signals at 65~Hz and 140~Hz, falling within the bandgaps of 1 and 2, respectively. 
The idea is to selectively suppress the low-frequency component 
in Region 1 and the high frequency one in Region 2 to establish different drop motion conditions between the regions. 
Now, if we consider a heterogeneous drop population mixing two drop sizes, since drops of different volumes resonate at different frequencies - and therefore tend to respond to substrates vibrating at different frequencies - we can expect the two regions to engage the two drop types differently. 

To test this idea, we deposit on each region two sets of four drops having different volumes (1.2 and 6.0~$\mu$L). Figure~\ref{fig:f4}(c) shows the response in Region 1, where the 65~Hz excitation is suppressed and the 140~Hz component dominates. Here, the larger drops feel the lower bandgap and remain stationary, and motion is limited to the smaller drops. We report a slight displacement of one large drop, attributable to its proximity to the actuation source. 
In contrast, in Region 2 (Fig.~\ref{fig:f4}(d)) the 140~Hz component is filtered out by the higher bandgap, resulting in stationarity of the smaller drops and activation of the larger drops only (Movie 7 in SM). Although the phase metric is not strictly defined for multi-frequency excitations, we can still evaluate the trajectory of points along the strip and 
predict accumulation points, shown in Figs.~\ref{fig:f4}(c-d). We appreciate that the type of drops that do move effectively cluster around these locations. This configuration practically works as a \textit{spatial volume filter}, which 
can selectively pick which size is allowed to move, cluster and segregate in a given region. While demonstrated here for two regions and two frequencies, this framework can be extended in principle to rainbow material substrates, 
heterogeneous drop populations and broad-band excitations, stretching the filtering capabilities.



\noindent 
The authors acknowledge support from National Science Foundation (grant CMMI-2211890) and are indebted to Z. Kujala and S. Lee for insightful discussions.

\bibliographystyle{apsrev4-1}
\bibliography{ref}

\end{document}


\preprint{APS/123-QED}

\title{Supplemental material: Centipede-Like Metastrip Enables On-Demand Programmable Drop Motion} 

\author{Sergio Britto}
\affiliation{%
Department of Civil, Environmental, and Geo- Engineering,
University of Minnesota, Minneapolis, MN 55455, US}

\author{Stefano Gonella}%
\email{sgonella@umn.edu}
\affiliation{%
Department of Civil, Environmental, and Geo- Engineering,
University of Minnesota, Minneapolis, MN 55455, US}%

\date{\today}
\maketitle

\section{IP and OOP response along the strip}

\begin{figure}
\includegraphics[width=\columnwidth]{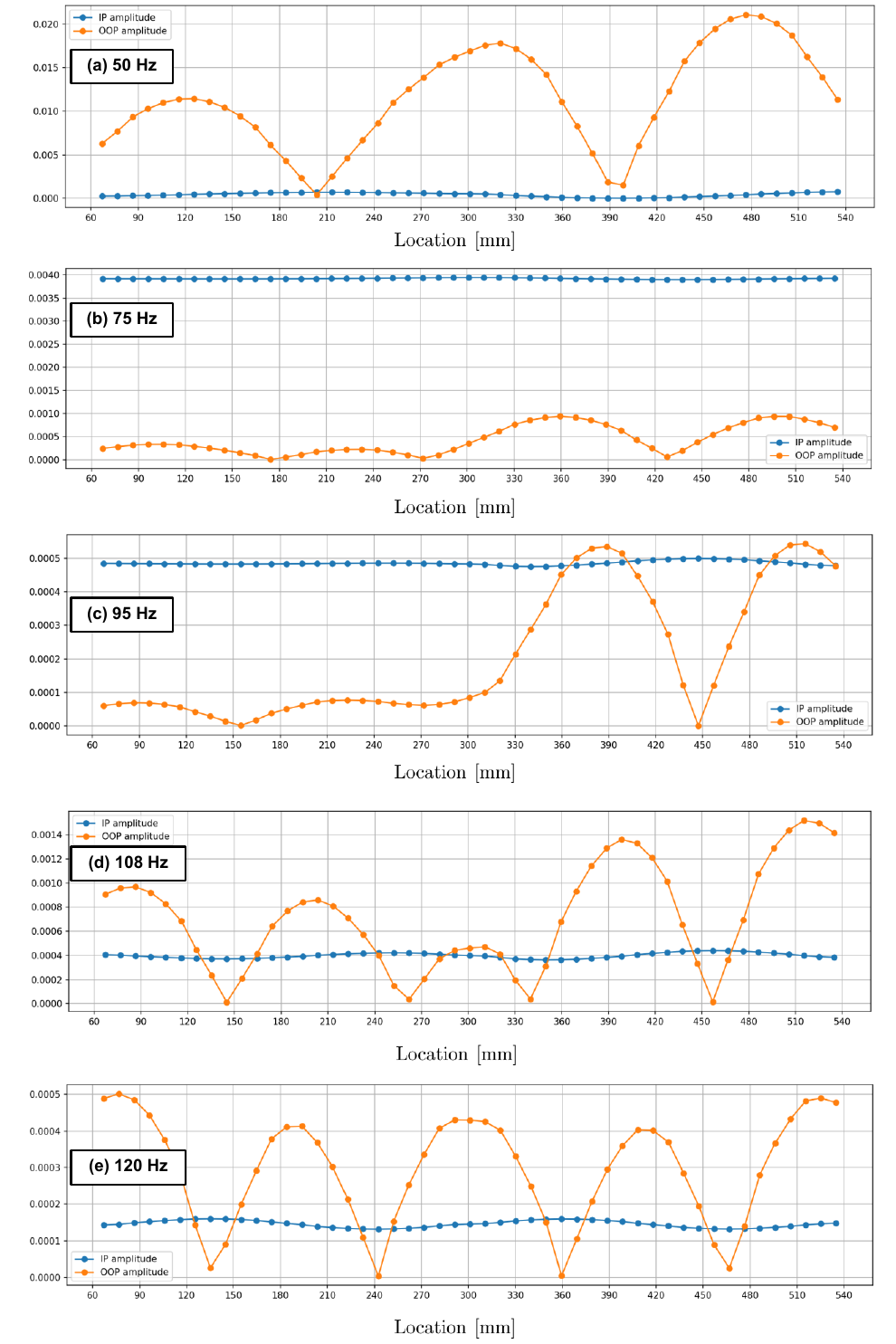}
\caption{\label{fig:f1_sm} 
Absolute amplitudes [mm] of in-plane (IP) (blue) and out-of-plane (OOP) (orange) for a few representative excitation frequencies (50, 75, 95, 108, and 120 Hz) along the span of the strip.
}
\end{figure}

Finite element (FE) simulations of the in-plane (IP) and out-of-plane (OOP) dynamics are conducted at representative excitation frequencies, shown in Fig.~\ref{fig:f1_sm}. The results demonstrate that, at a given frequency, the IP response is mostly spatially uniform and therefore could be considered effectively rigid, while all spatial heterogeneity in the displacement field along the strip arises from the OOP degree of freedom.

\section{Amplitude of excitation}

The excitation force applied through the stinger is regulated by the shaker input voltage, which is tuned to compensate for low-response modes of the structure and to avoid excessive vibration that could lead to structural damage near resonance peaks. For the pristine configuration, the applied shaker voltage ranged from 7.5 to 12 Vpp (peak-to-peak), with values of 10 V at 50 Hz, 12 V at 60 Hz, 10 V at 75 Hz, 7.5 V at 95 Hz, and 12 V at 105 Hz. For the second experiment (Region A with three additional tip masses), the input voltage was increased up to 15 Vpp to ensure that the droplets did not move within the bandgap region. For the final experiment (multi-frequency excitation), the input voltage was set to 9 Vpp. 

\section{Drop motion videos - Full Name List}

\begin{enumerate}
    \item Movie\_1\_Pristine\_48Hz 
    \item Movie\_2\_Pristine\_60Hz 
    \item Movie\_3\_Pristine\_75Hz 
    \item Movie\_4\_Pristine\_95Hz 
    \item Movie\_5\_Pristine\_105Hz 
    \item Movie\_6\_Bandgap\_110Hz 
    \item Movie\_7\_Multifrequency\_65\_140Hz
\end{enumerate}
